\documentclass[nofootinbib,twocolumn]{an}
\usepackage[fleqn]{amsmath}
\usepackage{graphicx}
\usepackage{times}

\begin{document}

\title{Constraints on supermassive black hole spins from observations\\ of active galaxy jets}

\author{E. Kun\inst{1}\thanks{Corresponding author:
  \email{kun@titan.physx.u-szeged.hu}},
 P. J. Wiita\inst{2},
 L. \'{A}. Gergely\inst{1,3},
 Z. Keresztes\inst{1},
 Gopal-Krishna\inst{4},
 P. L. Biermann\inst{5,6,7,8,9}
}
\titlerunning{Constraints on the supermassive black hole spin from observations on the jets}
\authorrunning{E. Kun et al.}
\institute{
Departments\ of Theoretical and Experimental Physics, University\ of
Szeged, Hungary
\and 
Department of Physics, The College of New Jersey, Ewing, NJ, USA
\and 
Department of Physics, Tokyo University of Science, Shinjuku-ku, Tokyo, Japan
\and
National Centre for Radio Astrophysics, TIFR, Pune, India
\and
Max Planck Institute for Radioastronomy, Bonn, Germany
\and
Institute for Nuclear Physics, Karlsruhe Institute for Technology, Karlsruhe, Germany
\and
Department of Physics \& Astronomy, University of Alabama, Tuscaloosa, AL, USA
\and
Department of Physics, University of Alabama at Huntsville, AL, USA
\and
Department of Physics \& Astronomy, University of Bonn, Germany}


\abstract{
We discuss the origin of the low-energy cutoff, or LEC, seen in the radio spectra of many extragalactic jets and relate this to the spin of the  supermassive black holes that presumably power them. Pion decay via proton-proton collisions is a possible  mechanism to supply a secondary positron population with a low energy limit. We expect that pion production would occur in advection dominated accretion flows or ADAFs. In radiatively inefficient ADAFs the heat energy of the accreting gas is unable to radiate in less than the accretion time and the particle temperature could be high enough so that thermal protons can yield such pion production.  Strong starbursts are another option for the injection of a truncated particle population into the jet. The role of both mechanisms is discussed with respect to the black hole spin estimate. The energy demanded to produce the pion decay process involves a minimum threshold for kinetic energy of the interacting protons. Therefore the mean proton speed in the flow can determine whether a LEC is generated.  In ADAFs the random velocity of the protons can exceed the minimum speed limit of pion production around the jet launching region in the innermost part of the flow. Finally we summarize the additional work needed to put the model assumptions on a more rigorous basis.
}

\maketitle

\section{Introduction}

It has been long argued that a high spin might be a necessary requirement
for some of the extremely powerful relativistic jets in AGN (e.g., Blandford \& Znajek 1977\nocite{Blandford1977}; Donea \& Biermann 1996\nocite{Donea1996}).
Although the specific formulae differ
substantially depending upon the jet launching model, the jet power in all
of them depends very sensitively on the BH spin (Meier 2003\nocite{Meier2003}; McKinney \& Narayan 2007\nocite{McKinney2007}; Tchekhovskoy, Narayan \& McKinney 2010\nocite{Tcheckhovskoy2010}).
The converse does not hold: an observed high jet power does not
necessarily imply a high spin. For such a conclusion it would be essential
to know the mass of the central BH and its associated magnetic
field at the base of the jet, which may or may not be maximal.

This begs the question: what other evidence is available for a high spin,
apart from modeling jets? There are a number of radio galaxies for which the
radio spectra suggest a low-energy cutoff or LEC in their energetic electron
spectra (Leahy, Muxlow \& Stephens 1989\nocite{Leahy1989}; Carilli et al. 1991\nocite{Carilli1991}; Celotti \& Fabian 1993\nocite{Celotti1993}; Duschl \& Lesch 1994\nocite{Duschl1994}).
An attractive physical mechanism for this is pion decay resulting from proton-proton
collisions (Biermann, Strom \& Falcke 1995\nocite{Biermann1995}; Gopal-Krishna, Biermann \& Wiita 2004\nocite{GopalKrishna2004}), although alternative suggestions
 have been made (Lesch, Schlickeiser \& Crusius 1988\nocite{Lesch1988}). Using hadronic interactions and
pion production in a hot disk (Mahadevan 1998\nocite{Mahadevan1998}) to provide positrons from
pion decay does require a relativistic temperature in the accretion disk
near the foot of the jet. Such a high temperature is possible only in a low
density disk, which is typical for an \textquotedblleft advection dominated
accretion flow\textquotedblright\ or ADAF. These ADAFs are commonly believed
to describe the low accretion rate regime (Narayan \& Yi 1994\nocite{Narayan1994}). The
extremely high temperature, in turn, seems to require a BH angular momentum
parameter above 0.95 (Falcke \& Biermann 1995\nocite{Falcke1995}; Gopal-Krishna et al. 2004; Falcke, Malkan \& Biermann 1995\nocite{FalckeMalkan1995}). 

The mass of the pion created in proton hadronic interactions gives the
minimum Lorentz factor of a final decay product, an electron or positron 
(Gaisser 1991\nocite{Gaisser1991}; Stanev 2010\nocite{Stanev2010}). Thus the leptons emitting in radio galaxies might be predominantly
secondary, having this low energy limit (Gopal-Krishna et al. 2004). As the 511 keV emission in the
Galactic Center indicates, there can also be large positron production as a
consequence of prodigious star formation and the ensuing supernova
explosions (e.g., Biermann et al. 2010\nocite{Biermann2010}). The difference in radio spectra is in the temporal variability of a core; however,  supernova
explosions also show temporal variability in their non-thermal radio emission,
but no relativistic motion except in the rare cases of a GRB.

A strong starburst may provide a possible alternative to explain a low
energy cutoff in the cosmic ray lepton spectrum, although it obviously will
never produce the brightness temperature of a compact relativistic jet. In
the observed cosmic rays, the positron fraction is known to rise towards
lower energies (Protheroe 1982\nocite{Protheroe1982}; Adriani et al. 2009\nocite{Adriani2009}), which can
be explained as interactions in the environment of massive stars (Biermann et al. 1991\nocite{Biermannetal1991}; Biermann 1998\nocite{BiermannNuAsproc1998}; Biermann et al. 2009\nocite{BiermannBecker2009}; Biermann et al. 2010\nocite{Biermann2010};  Biermann et al. 2012\nocite{Biermann2012});
the model prediction is that the positron fraction increases as $E^{-5/9}$
toward lower energies. It is important to realize that such a model can only
be verified with isotope ratios, since the influence of the Solar modulation
can then be minimized. Normal Solar modulation affects all charged particles
with energies ~ \lower2pt \hbox {$\buildrel <
\over {\scriptstyle \sim }$}~ 10 GeV per unit charge, and cuts off all
particles below near 300 MeV per unit charge, as measured outside the Solar
system. However, using isotope ratios, the above energy dependence has been
verified and found to be a very good fit to
the data (Ptuskin et al. 1999\nocite{Ptuskin1999}). Extrapolating this energy dependence, we find that near the
production threshold of about 50 MeV the positrons approach half of all
leptons. 
This is valid in our Solar neighbourhood, and if we consider more active
regions, with more interstellar matter, it is plausible to assume that some
regions will have a positron flux at such an energy actually exceeding the
primary cosmic ray electron flux; obviously, the secondary electron flux
from $\pi ^{\pm }$ decay also shows the LEC. If this were true, then the
overall flux of electrons and positrons --- being mostly secondary at
50--100 MeV --- would show the same LEC as in the other scenario, where they
are produced by hadronic interaction in a hot accretion disk (Mahadevan 1998).

\section{SMBH spin and LEC}

Thus we have two clear options to generate a LEC in the relevant lepton
spectra, though other possibilities we have not discussed may also exist,
such as from a hypothetical dark matter particle decay. One is a hot,
radiatively inefficient, accretion disk at the base of the jet. The other is
a starburst near the central region of the host galaxy. Both sites can
inject an electron energy distribution with a LEC. In the first case, the
spin has to be very high to allow such a hot accretion disk. In the second
case, we require a strong starburst in the central region of the host
galaxy, such as is present in Centaurus A, for instance, with a large
luminosity in the far-infrared from the massive stars irradiating the
surrounding dust (Popescu et al. 2002\nocite{Popescu2002}; Pierini et al. 2003\nocite{Pierini2003}).

It follows, that in a case where we do observe a LEC in the electron energy
distribution through a low frequency cutoff in the radio spectrum, both of
the above explanations should be considered. If the LEC is present despite
no sign of any obvious starburst, a large SMBH spin is likely to be
required. However, if a strong starburst is observed, this would
obviate the need for a central BH with a high spin. 
A spinning SMBH with a smaller merging companion at the base of the jet was also investigated (T\'{a}pai et al. in prep.\nocite{Tapai2012}).

\section{Pion production in the ADAF}

\subsection{Proton energy: a key parameter}
In proton-proton collisions the partial cross-section rises to a high peak for pion production, far above the partial cross-section for electron-positron pair production, and therefore we focus on pion decay electrons and positrons, which have a relatively high energy. At relativistic particle energies the collision of two protons typically yields a proton, a neutron and a positively charged pion. Pions decay further according to the process $\pi^+ \rightarrow$ $e^+$ $+ \nu_e + \overline{\nu_\mu} + \nu_\mu$, where the secondary positron population  shows the LEC.  At lower energies some hydrogen fusion occurs and as a consequence a primary positron population is formed. We are interested in the first case. A threshold energy for pion production comes from the conservation of
energy, assuming two protons moving with equal and opposite velocities
(then the total momentum is zero):
\begin{eqnarray}
E=2\gamma_\mathrm{r} m_{\mathrm{0,p}}c^2=m_{\mathrm{0,p}}c^{2}+m_{\mathrm{0,n}}c^{2}+m_{\mathrm{0,\pi ^{+}}}c^{2}. 
\end{eqnarray}
Substituting the rest mass of particles the above equality yields
$\gamma_{\mathrm{r}}=1.08$
and $v_{\mathrm{r}}=0.37c$ 
in the relativistic frame. Using the relativistic velocity addition formula, it corresponds to $v_{\mathrm{p, min}}= 0.65$\,c and $\gamma_{\mathrm{p,min}}=1.31$
 in the rest frame of one proton.
Then a required minimum kinetic energy for the protons: $E_{\mathrm{kin}}=\gamma E_{\mathrm{m_{0,p}}}-E_{\mathrm{m_{0,p}}}\approx 290$\,MeV.
We expect that the presence of such high proton kinetic energies would be a good indicator of an environment adequate to produce substantial pion production. 

\subsection{Advection Dominated Accretion Flows}

A unified description of black hole accretion disks supplies four equilibrium solutions for the differentially rotating viscous flows (Chen et al. 1995\nocite{Chen1995}). In the ADAF the heat energy is stored as entropy of the accreting gas, which leads to radiative inefficiency $\eta \equiv L/\dot{M}c^2\ll 0.1$ (Narayan \& McClintock 2008)\nocite{Narayan2008}.

The ADAF solutions are characterized by extremely low density, large pressure and sub-Keplerian rotation. 
The gas has positive Bernoulli constant, which means it is not bounded to the central BH. Therefore if the orbiting matter reverses its direction in the flow it would reach infinity with a positive energy.
This suggests that an ADAF likely generates powerful winds and relativistic jets. The observed radio and X-ray radiations of some low-luminosity AGNs and Fanaroff-Riley I type radio galaxies are explainable by using of a coupled jet plus an underlying ADAF (e. g., Yuan, Markhoff \& Falcke 2002\nocite{Yuan2002}; Yuan et al. 2002\nocite{Yuan2002}), which again suggests a connection between the jets and the ADAFs.

The  gas density $\rho$, its radial and angular velocities $v$ and $\Omega$ and the isothermal sound speed $c_s$ change in the flow according to the differential equations presented below  (e.g.,  Narayan \& Yi 1994\nocite{Narayan1994}). They describe a two dimensional, steady state, axisymmetric flow. The equations are vertically averaged and all variables of the flow motion have dependence just on their $R$ coordinate ($\partial /\partial \phi =0,\partial /\partial t=0$ in the $R\phi$ plane; $\phi$ is the azimuthal coordinate). The continuity equation, radial and azimuthal components of the momentum equation and the energy equation, respectively are:
\begin{flalign}
& \frac{d}{dR}(\rho RHv)=0,\\
& v\frac{dv}{dR}-\Omega ^{2}R=-\Omega _{K}^{2}R-\frac{1}{\rho }\frac{d}{dR}%
	(\rho c_{s}^{2}),\\
& v\frac{d(\Omega R^{2})}{dR}=\frac{1}{\rho RH}\frac{d}{dR} \left( \frac{\alpha
	\rho c_{s}^{2}R^{3}H}{\Omega _{K}}\frac{d\Omega }{dR} \right),\\
& \Sigma vT\frac{ds}{dR}\!=\!\frac{3\!+\!3\epsilon }{2}2\rho Hv\frac{dc_{s}^{2}}{dR}%
	\!-\!2c_{s}^{2}Hv\frac{d\rho }{dR}\!=\!Q^{+}\!-\!Q^{-}. 
\end{flalign}
Here $Q^+$ measures the energy input per unit area (due to viscous dissipation) and $Q^-$ is the outward flowing energy per unit area (due to radiative cooling). The left hand side of the energy equation (5) is the advected entropy and the energy equation in a compact form is $Q^{adv}\!=\!Q^+-Q^-$.
As shown by Narayan and Yi (1994) these equations have self similar solutions in the form of scaling laws:
\begin{flalign}
& v=-(5+2\epsilon')\frac{g(\alpha ,\epsilon')}{3\alpha}%
	v_{K},\\
& \Omega =\left[ \frac{2\epsilon'(5+2\epsilon')g(\alpha
	,\epsilon')}{9\alpha ^{2}}\right] ^{1/2}\Omega _{K},\\
& c_{s}^{2}=\frac{2(5+2\epsilon')g(\alpha
	,\epsilon')}{9\alpha ^{2}}v_{K}^{2},\\
& g(\alpha ,\epsilon')\equiv \left[ 1+\frac{18\alpha ^{2}}{%
	(5+2\epsilon')^{2}}\right] ^{1/2}-1,
\end{flalign}
where $\epsilon\!=\!(5/3-\gamma)/(\gamma-1)$ is a parameter of the flow ($\gamma$ is the ratio of specific heats, $\gamma\!=\!5/3\rightarrow \epsilon\!=\!0,\gamma\!=\!4/3\rightarrow \epsilon =1$). The Keplerian velocity and angular velocity are denoted as $v_K\!=\!(Gm/R)^{1/2}$ and $\Omega_K\!=\!(Gm/R^3)^{1/2}$, respectively, where G is the gravitational constant and $m$ is the central mass.  Let $f$ measure the degree of advection so that if $f\!=\!1$, the total amount of the viscous heat is stored in the particles and this limit yields an extreme ADAF ($Q^-=0$). Whereas with f = 0 the radiative cooling is very effective and this limit yields an extreme thin disk solution. $\epsilon'\!\equiv\!\epsilon/f$. All of these quantities are now expressed solely in terms of $R$, $\alpha$ and $\epsilon'$; the last two of these parameters are suitable for a description of the nature of the flow.
\subsection{Speed limit around the jet launching region}
Since $\alpha^2\!\ll\!1$ in ADAFs we can expand $g(\alpha ,\epsilon')$ in powers of $\alpha$ and this yields
\begin{eqnarray}
	\label{cs2vk}
	c_{s}^{2} \approx \frac{2}{5+2\epsilon'}v_{K}^{2}.
\end{eqnarray}
As the protons in the plasma are not significantly relativistic, we first assume that protons can be treated as an ideal gas to a decent approximation and particle speed is characterized by the Maxwell-Boltzmann distribution. Then the adiabatic index should be quite close to $5/3$. According to the equation of state of an ideal gas, the isothermal sound speed is expressible as a function of $T$,
\begin{eqnarray}
\label{cs2}
	c_s^2=\frac{P}{\rho} \rightarrow c_{s}^{2}=\frac{k T}{\overline{\mu}m_H(=\overline{m})},
\end{eqnarray}
in a gas pressure dominated accretion flow and  if we assume a pure hydrogen plasma $\overline{m}=(m_p+m_n+m_e)/3$. Eq. (10) and Eq. (11) yield a $T(\epsilon',R,m$) function:
\begin{eqnarray}
	T\approx \frac{\overline{m}}{k} \left[ \left( \frac{2}{5+2\epsilon'} \right)\frac{Gm}{R}\right].
\end{eqnarray}
The radial velocity in an ADAF disk is usually high compared to that of a standard thin disk,  since, while $\alpha$ is low, $v \sim \alpha c_s^2 / v_K$ and $c_s$ is large. Therefore the random velocities of protons in the disk could be sufficiently high to reach the minimum speed requirement for pion production in the relativistic temperature regime of the ADAF, since random motions in the disk are related to its temperature.
 According to the Maxwell-Boltzmann (MB) distribution of the particle speeds we can calculate the probability (P) that the proton speed is between the required minimum value of the workable proton collisions  $v_{\mathrm{p,min}}\!\approx\!0.65$\,c (which provide a pion) and $v_{\mathrm{p,max}}\!\approx\!1$\,c at a given temperature or radius. 
\begin{figure}[tbp]
\begin{center}
\includegraphics[width=6cm]{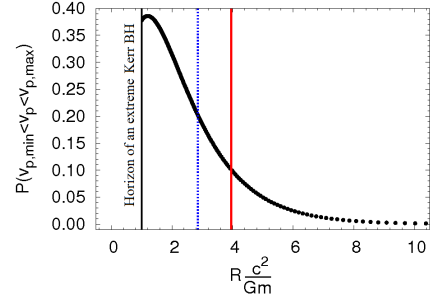}
\end{center}
\caption{Probability P of pion production inside of a radiatively inefficient ADAF. R denotes the radial coordinate of the accreting particle in the R$\phi$ plane. Red (continuous) and blue (short dashed) lines imply the 0.2 and 0.1 levels of the probability, respectively. Plasma is gas dominated and $\gamma\!=\!5/3$, then $\epsilon'\!=\!0$.}
\label{fig1}
\end{figure}

In Fig.\ 1 all points arise from a numerical integration of the MB
distribution between the required minimum speed and the maximum speed at a given
temperature. In our oversimplified model strong pion production will only occur near to the black hole and as a consequence a LEC showing positron population can be injected into the jet.  However, in this regime general relativity effects are strong and the scaling laws break down (Gammie \& Popham 1998\nocite{Gammie1998}).

 On the other hand, if the flow is not highly advection dominated then there is a significant contribution from radiation pressure and the local pressure in the disk is the sum of the pressure of matter and radiation $P\!=\!P_r\!+\!P_m$. $P_r\!=\!(\gamma\!-\!1) \varepsilon_r$ and if the photons are in equilibrium then $\gamma\!=\!4/3$ and $P_r\!=\!\frac{1}{3}aT^4$ where $a\!=\!4\sigma/c \approx 7.56 \cdot 10^{-16}\mathrm{J\,m^{-3}\,K^{-4}}$.
 Under these circumstances the simple scaling law relations for the quantities in the ADAF fail as the ratio of the thermal to the radiation pressure depends on $R$. Then the effective value of $\gamma$ will vary between $5/3$ when fully gas pressure dominated down toward $4/3$ when the radiation pressure is much greater. In that case the general expression $c_s^2\!=\!dP/d\rho$ must be used and one must obtain the solution numerically.
While we have assumed the protons have a thermal distribution, there is an indication that for all accretion rates the Coulomb collisions may be too inefficient to thermalize the protons (Mahadevan \& Quataert 1997)\nocite{Mahadevan1997}.

\section{Concluding Remarks}

The supermassive black holes at the centers of galaxies and their immediate environs can emit a pair of relativistic jets.  The morphology and energetics of these relativistic jets are tightly
coupled to the central engine and a careful analysis can provide
interesting possibilities to understand the SMBH behaviour.

In this paper we established an explanation of the low-energy cutoff seen in the synchrotron spectra of many radio active galaxies and operating under the reasonable assumption that the LEC arises through pion decay processes via protonic collisions. 
Only a low density and very hot ADAF is able to maintain the relativistic temperature which is required to yield substantial pion production. Such a very hot disk requires the dimensionless spin parameter $a\!>\!0.95$.
Another possible hypothesis that could inject secondary po\-sitrons into the jet would be a strong starburst being present in the central region of the host galaxy. That situation would not require a high black hole spin, but in the
absence of a starburst the existence of a LEC indicates a high spin of
the central black hole. 
We have calculated the minimum kinetic energy such that colliding
protons generate the desired decay products. The speed of the colliding protons must exceed  $v_{p,min}\!\approx\!0.65$ c to allow for pion production.

Since an ADAF is the only type of accretion flow that seems capable of producing such high kinetic energies, we summarized those differential equations and their self similar solutions. We connected the ADAF temperature to the typical random velocities of protons trough the isothermal sound speed. In a simplified model, we treated protons as an ideal gas in thermal equilibrium and assumed that the flow is highly advection dominated. We have shown that below about $2$ Schwarzschild radii the random speed of $10$ percent of protons can exceed the ``speed limit'' and so around the innermost region the pion production via proton collisions should be an important process. Pions are unstable, they will decay further to positrons and neutrinos, and so that positron population will be imprinted with a LEC. Future work will include the analysis of the general case where the plasma is predominantly thermal but radiation pressure plays a significant role, as well as an analysis for the case where the protons do not come into thermal equilibrium. A general relativistic treatment and use of Maxwell-J{\"u}ttner distribution to determine P would be another key extensions.
\acknowledgements
L\'{A}G was supported by EU grant T\'{A}MOP-4.2.2.A-11/1/KONV-2012-0060 and the Japan Society for the Promotion of Science; ZK by OTKA grant 100216.

\end{document}